\documentclass[10pt]{article}
\begin{document}

\newcommand{\m}[1]{\ensuremath\mbox{\boldmath $#1$}}
\newcommand{\be}{\begin{equation}} \newcommand{\ee}{\end{equation}}
\newcommand{\ba}{\begin{eqnarray}} \newcommand{\ea}{\end{eqnarray}}
\newcommand{\nn}{\nonumber} \renewcommand{\bf}{\textbf}
\newcommand{\ra}{\rightarrow} \renewcommand{\c}{\cdot}
\renewcommand{\d}{\mathrm{d}} \newcommand{\diag}{\mathrm{diag}}
\renewcommand{\dim}{\mathrm{dim}} \newcommand{\D}{\mathrm{D}}
\newcommand{\integer}{\mathrm{integer}}\newcommand{\LL}{\lambda}
\newcommand{\R}{\mathbf{R}} \renewcommand{\t}{\mathrm{t}}
\newcommand{\T}{\mathbf{T}} \newcommand{\V}{\mathbf{V}}
\newcommand{\tr}{\mathrm{tr}} \newcommand{\cA}{\cal A}
\newcommand{\cB}{\cal B} \newcommand{\cC}{\cal C}
\newcommand{\cD}{\mathrm{\cal D}} \newcommand{\cF}{\cal F}
\newcommand{\cG}{\cal G} \newcommand{\cL}{\cal L}
\newcommand{\cO}{\cal O} \newcommand{\cT}{\cal T}
\newcommand{\cU}{\cal U} \newcommand{\s}{\,\,\,}
\renewcommand{\a}{\alpha} \renewcommand{\b}{\beta}
\newcommand{\e}{\mathrm{e}} \newcommand{\eps}{\epsilon}
\newcommand{\f}{\phi} \newcommand{\fr}{\frac} \newcommand{\g}{\gamma}
\newcommand{\h}{\hat} \renewcommand{\i}{\mathrm{i}}
\newcommand{\p}{\partial} \newcommand{\w}{\wedge} \newcommand{\x}{\xi}
\newcommand{\NN}{\vec \nabla}

\input{epsf}

\title{The Bohr Atom of Glueballs}

\author{John P. Ralston}

\maketitle

  \medskip
 
\begin{center}
    
Department of Physics \& Astronomy, University of Kansas,\\
Lawrence, KS 66045\\

\end{center}

\begin{figure}
	
\epsfxsize=3in \epsfysize=2.8in
  
\epsfbox{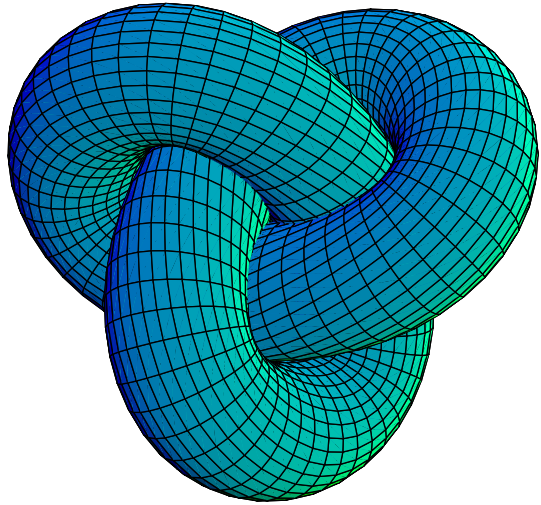}

\caption{A tightly wrapped trefoil knot, identified as the second
member of the glueball spectrum.}

\label{fig:trefoil}
\end{figure}

{\bf {Recently Buniy and Kephart\cite{BK} made an astonishing
empirical observation, which anyone can reproduce at home.  Measure
the {\it lengths} of closed knots tied from ordinary rope.  The
``double do-nut'', and the beautiful trefoil knot (Fig. 
\ref{fig:trefoil}) are examples.  Tie the knots tightly, and glue or
splice the tails into a seamless unity.  Compare two knots with
corresponding members of the mysterious particle states known as
``glueball'' candidates in the literature\cite{pdg}.  Propose that the
microscopic glueball mass ought to be proportional to the macroscopic
mass of the corresponding knot.  Fit two parameters, then {\it
predict} 12 of 12 remaining glueball masses with extraordinary
accuracy, knot by knot.  Here we relate these observations to the
fundamental gauge theory of gluons, by recognizing a hidden gauge
symmetry bent into the knots.  As a result the existence and
importance of a gluon mass parameter is clarified.  Paradoxically
forbidden by the usual framework\cite{pokorski}, the gluon mass cannot
be expressed in the usual coordinates, but has a natural meaning in
the geometry of knots.}}

The Buniy-Kephart (BK) discovery is dramatic and can be called the
``Bohr atom'' of glueballs.  Bohr's quantization\cite{bohr} is
explained by a whole number of vibrations of an electronic wave
function traversing a closed orbit.  The BK explanation postulates a
``solitonic'' magnetic flux rope of gauge fields, traversing a closed
but knotted path with a whole number of self- crossings.  The energy, 
and then the mass of the flux rope is proportional to the length of the 
rope. The glueball mass spectrum follows\cite{BK}.

Deep questions of consistency hide in this picture.  The fundamental
theory of glue\cite{pokorski}, Quantum Chromodynamics (QCD), predicts
glueballs\cite{gluestuff} only indirectly, through the existence of
certain singlet operators.  Decades have passed seeking a clear
signature\cite{gluestuff}.  The QCD static energy density has a term
going like $\frac{1}{2}\vec B^{2}$, the magnetic energy density. 
Ordinary magnetic flux (the field lines of ``bar-magnets'') flows in
closed loops, yet strict flux conservation is {\it not} a general
property of the more elaborate chomo-flux of QCD\cite{pokorski}.  The
hopeful logic holds that {\it if} a flux is conserved and arranged
into a constant width tube, {\it then} the classical energy and
glueball mass goes like the length of the knot.  So far so good: yet
the theory has no solitons!  QCD and other {\it gauge theories} lack a
mass scale upon which to base any particular soliton mass.  The
quantum treatment inducing a scale called $\Lambda_{QCD}$ does not
change this.  Moreover, the requirements of a gauge theory are
exacting.  It is commonly held impossible to add a mass scale
affecting the infrared (large-scale) structure, and retain gauge
invariance, the {\it raison d'$\hat e$tre} of glue itself.

The culprit is {\it confinement}, the phenomenon that gauge fields and
quarks cannot get outside of the strongly interacting particles. 
Confinement is poorly understood.  ``Effective'' theories are proposed
as surrogates for the fundamental one.  Fadeev and Niemi\cite{fadeev}
constructed knotted solitons, such as the trefoil\cite{knot} (Fig. 
\ref{fig:trefoil}), in an ad-hoc effective theory.  Yet the picture of
conserved flux and knotted rope is a hybrid.  There has been no
direct connection between solitons, knots, and any underlying gauge
fields which form the fundamental glue.
 
Look afresh at the effective theory making knots.  The basic variable
is a real-valued 3-component unit vector field $\hat n(\vec x)$.  The
Lagrangian density\footnote{The action $S =\int d^{4}x L$. Units are
$\hbar=c=1$.  } is \ba L &= &\frac{m^{2}}{4}\p_{\mu} \hat n \cdot
\p^{\mu} \hat n - \frac{1}{g^{2}} (\hat n \cdot\p_{\mu} \hat n \times
\p_{ \nu} \hat n ) (\hat n \cdot \p^{ \mu} \hat n \times \p^{ \nu}
\hat n ) \nn \\ & =&m^{2}L_{2}+ \frac{1}{g^{2}} L_{4}. 
\label{nlagrange}\ea Here $\p_{\mu}=\p/\p x^{\mu}$, while $\cdot$ and
$\times $ denote the dot and cross product of three dimensional space. 
No flux tubes are obvious in Eq.  \ref{nlagrange}.  Nor are local
transformations of $\hat n(x)$  a symmetry.  Therefore if the
theory is related to a gauge theory, we propose it is the {\it
invariant} coordinatization of a gauge theory.

An invariant formulation is possible by embedding gauge-theory
geometry in a larger space.  Interpret $\hat n(x)$ as a vector
perpendicular to a 2-surface, spanned by a local tangent frame $\hat
e^{a}$, $a=1, \, 2$, $\hat e^{a} \cdot \hat e^{b}=\delta^{ab} $. 
Transfer attention to the surface.  Its bending and stretching fixes
the system's energy. Surface coordinates are related non-invertibly by
$$\hat n =\hat e^{1} \times \hat e^{2}.$$ Compare the freedoms
of the $\hat n, \, \hat e$ descriptions: use of the tangent-frame ``inner''
$\hat e$'s involves one extra angle $\phi(x)$. This angle parametrizes the
orientation of the frame on the surface.  Angle $\phi(x)$.is not determined
by the Lagrange density depending on $\hat n(x)$ and can be freely
chosen as an arbitrary smooth function of $\vec x$.  There is a {\it
local} symmetry \ba \left(
\begin{array}{c } \hat e^{1}(x) \\ \hat e^{1}(x) \end{array}
\right) &\ra &R(x) \left(
\begin{array}{c } \hat e^{1}(x) \\ \hat e^{1}(x) \end{array}
\right) = \left( \begin{array}{cc } cos\phi(x) & sin \phi(x) \\
   -sin\phi(x) & cos\phi(x) \end{array} \right) \left(
\begin{array}{c } \hat e^{1}(x) \\ \hat e^{1}(x) \end{array}
\right); \nn \\ \hat n(x) &\ra &\hat n(x) .  \label{gaugerot} \ea Due
to local invariance of $\hat n$, the system dynamics has a local
$S0(2)$ gauge symmetry when expressed via the $e$'s.  This happens to
be just the same symmetry upon which flux tubes are based.

Let us explore the meaning of the separate terms.  Some algebra yields
\ba \hat n \cdot \p_{\mu} \hat n \times \p_{\nu} \hat n =-\frac{
1}{2}( \, \p_{\mu} \hat e_{k}^{1} \p_{\nu} \hat e_{k}^{2} -\p_{\nu}
\hat e_{k}^{1} \p_{\mu} \hat e_{k}^{2} \,)\label{curl} .\ea A famous
theorem says that invariants of local transformations must be a
function of gauge-covariant derivatives\cite{utiyama}.  Differential
geometry defines a {\it connection} $A_{\mu} = \frac{1}{g } \hat e_{k}
^{1} \p _{\mu} \hat e^{2}$ to be used.  Under Eq.  \ref{gaugerot} we
have \ba A _{\mu} \ra A _{\mu}' (R(x) e) =A_{\mu} (e(x))+ \p_{\mu
}\phi(x) , \ea following by definition, and $A_{\mu} $ serves as a
{\it gauge field}.  Very nicely, \ba \frac{1}{g}\hat n \cdot \p_{\mu}
\hat n \times \p_{\nu} \hat n & = & -\frac{1}{2}( \,
\p_{\mu}A_{\nu}-\p_{\nu}A_{\mu} \,), \nn \label{fdefined} \\ & \equiv
&-\frac{1}{2} F_{\mu \nu }; \\
\frac{1}{g^{2} } L_{4}& = & -\frac{1}{4}F_{\mu \nu }F^{\mu \nu } \label{L4defined}.\ea
We find that $L_{4}$ actually is the usual Lagrangian of a hidden
gauge theory!  {\it Flux conservation is established,} defining $
B_{i} =\frac{1}{2} \epsilon_{ijk}F_{jk}$, with $\vec \nabla \cdot \vec
B=0$, a Bianchi identity, being the ancient law that ``you can't break
a magnetic rope''.

What is the meaning of the $L_{2}$ term?  Algebra gives \ba
\frac{m^{2}}{4}\p_{\mu} \hat n \cdot \p^{\mu} \hat n =
-\frac{m^{2}}{2}(A_{\mu}A^{\mu}-\frac{1}{2}\p_{\mu}\hat e_{k} ^{a}
\p_{\mu}\hat e_{k} ^{a} ); \nn \\ L=-\frac{1}{4}F_{\mu \nu }F^{\mu \nu
} -\frac{1}{2}m^{2}A_{\mu}A^{\mu}+\frac{1}{4}m^{2} \p_{\mu}\hat e_{k}
^{a} \p_{\mu}\hat e_{k} ^{a} .  \label{alagrangian} \ea Geometry
proves it is impossible to express $L$ entirely as a local function of
$A_{\mu}.$ The geometrical meaning of $L_{2}$ is the sum of the
squares of the principal curvatures of the bent and stretched
2-surface.  The {\it extrinsic} (``bending'') curvatures depend on the
embedding of the 2-surface in a higher space.  In contrast, only {\it
intrinsic} curvatures independent of embedding are expressed by $A$.  

Dynamically, parameter $m$ defines an {\it effective gluon mass}. 
Addition of $(\p_{\mu}e)^{2}$ terms gives Eq.  \ref {alagrangian} a
different mass from the usual, non-gauge-invariant kind.  Recall that
varying $L_{4}$ with respect to $A$ would give the Yang-Mills
(Maxwell) equations in the usual gauge theory.  Instead vary the
action with respect to frames $\hat e^{a} $, which after fixing the
gauge, are just the same as varying with respect to $\hat n$.  There
are extra solutions because the bending of the knot has real physical
energy in all forms of the knot's curvature.

Does the same pattern extend to the non-Abelian theory?  The answer is
{\it yes}.  Make incomplete frames $e_{K}^{a}$ on $ K=1\ldots K_{max}$
complex dimensions.  For a unitary group the frames are orthonormal,
$e_{K}^{a}\bar e_{K}^{b}=\delta^{ab}, \, a,\, b=1\ldots N. $ Let the
frames transform as fundamental representations of a local group
$U(x)$, $e_{K}^{a}\ra U^{ab}(x) e_{K}^{b}$.  The induced gauge field
is $A^{ab}_{\mu} = -\frac{i}{g} e_{K}^{a}\p_{\mu}\bar e_{K}^{b}, $
with bar denoting the complex conjugate and $g$ the coupling constant. 
The gauge field transforms as $A(e, \, \bar e ;\, x) \ra A(Ue,\, \bar
e U^{-1 } ; \, x) =U(x)A(x)U^{-1}(x)-\frac{i }{g} U(x) \p U^{-1}(x) ,
$ with indices suppressed.  To allow field strength $F_{\mu \nu}^{ab}
\neq 0 $, the frames must be embedded in a space of dimension larger
than the one they span: $K_{max} >N$.  The formula for $m^{2 }
L_{2}=-\frac{m^{2 } }{2}tr( A^{2 }-\p e \p \bar e)$, using ``bar'' for
complex conjugation and $tr$ for trace over the indices, is {\it
unique} and describes the lowest-order invariant.

 Now ask again: How can it be posible that the modified gauge theory,
 with its gauge invariance and conserved magnetic flux, might have
 soliton masses proportional to the knot-lengths?  The energy density
 from Eq.  \ref{alagrangian} consists of two terms, $m^{2}h_{2}$ and
 $\frac{1}{g^{2} }h_{4}$ with 2 and 4 derivatives, respectively. 
 Suppose we find a static solution $\hat n(\vec x)$.  Compare its
 energy $E$ with the energy $E(\lambda)$ of a re-scaled configuration
 $\hat n_{\lambda } (\vec x)= \hat n( \lambda \vec x)$.  Change
 variables to integrate over $\vec x_{\lambda}= \lambda \vec x $. 
 This gives \ba E(\lambda) = \int \frac{d^{3}
 x_{\lambda}}{\lambda^{3}}\, \lambda^{2} m^{2} h_{2}(\hat n_{\lambda
 }(x_{\lambda} ))+\frac{1}{g^{2}} \lambda^{4} h_{4}(\hat
 n_{\lambda}(x_{\lambda})).  \ea The energy $E(\lambda)$ is stationary
 for all variations. Varying $\lambda$ at $\lambda=1$ must be
 stationary.  This yields \ba \int d^{3} x \,
 m^{2} h_{2}(\lambda=1) & =& \int d^{3} x \,\frac{1}{g^{2}}
 h_{4}(\lambda=1) \equiv E_{4} ; \nn \\
\:\:\:\:thus\: \: \nn E(\lambda=1) &=& 2 E_{4} \ea Using Eqs.  \ref
{fdefined}, \ref{L4defined} the energy $E_{4}$ is the magnetic energy
density cited earlier.  The knotted soliton mass \ba
M_{knot}=\frac{2}{2}\int d^{3}x \, \vec B^{2},\ea {\it and the knotted
soliton mass is proportional to the knot volume}, just as proposed by
BK. To complete the chain of logic, knot-volumes must go like the {\it
lengths} of knots, implying constant rope width.  This was already
shown\cite{fadeev}, although not yet shown for {\it all} knots. 
Industrially making higher order soliton knots is itself mind-bogglng
in terms of variable $\hat n$.  We suggest a procedure: First bend a
solenoid along the knot.  Solve a trial $\vec A$ with the right
topology.  Settle into the appropriate soliton by using a numerical
relaxation method.

\begin{figure}
	
\epsfxsize=4.5in \epsfysize=2.5in
  
\epsfbox{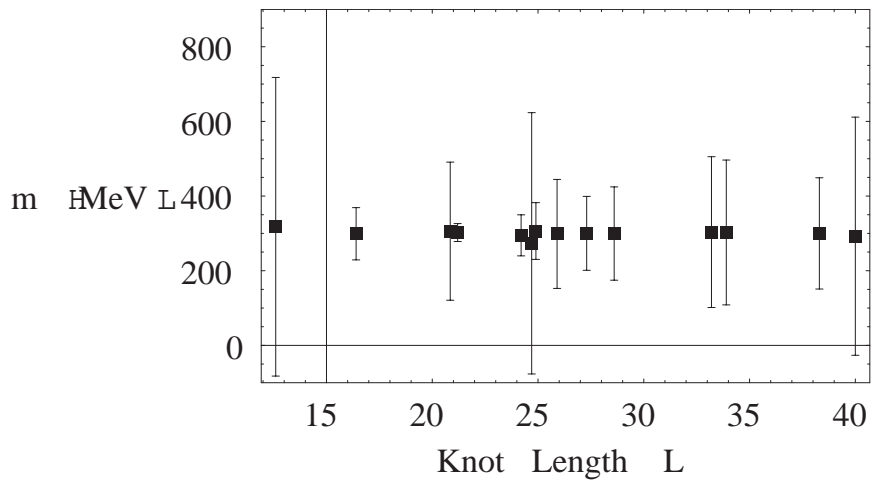}

\caption{ Gluon mass fitting parameters $m(j)=\frac{M_{j}-\Delta M
}{\beta L(j) },$ as a function of knot length $L(j)$.  Deduced mass
values $m_{j}$ are nearly constant (dashed line $m=298$ MeV),
indicating one universal mass parameter $m$ fits the data.  The
overall scale from $\beta$ is arbitrary.  Error bars are the
experimental widths of the state.  Error bars far exceeding the
fluctuations of values is not a statistically statifactory pattern of
data.  Error bars are about 10 times smaller using the experimental
errors of the state mass parameters.  Then the data's behavior becomes
statistically unremarkable, but physically inexplicable.  }

\label{fig:glueplot}
\end{figure}

The theory of Eq.  \ref{alagrangian} is superbly suited to the
phenomenological observations of Ref.  \cite{BK}.  To reiterate this
conclusion, the data for the masses of the glueballs is inverted to
find the gluon mass value.  This restates observations in Ref. 
\cite{BK}, and is not an independent test.  Soliton masses scale like
$m$, the gluon mass parameter, as the sole scale.  For each glueball
candidate mass $M_{j}$ we then calculate $m_{j}$, a trial mass
parameter.  The relation is \ba m(j)= \frac{ M_{j}-\Delta M }{\b
L(j)}.\ea Here $\Delta M$ is a free parameter, hopefully small,
representing quantum corrections.  Take $L(j)$ from knot theory, made
dimensionful with parameter $\beta$, which absorbs $g^{2}$ and the
knot width-to-length ratio.  The idea fails if the $m(J)$ take many
different values.  But the mass parameters $m(j)$ (Fig. 
\ref{fig:glueplot}) are found remarkably constant.  One universal
gluon mass: $m(j) \ra m=$298 $\pm$ 19 MeV, $\Delta M=15.0 \pm $84 MeV
is supported by the fit.  Parameter $\beta$ is not determined, and was
adjusted so that the gluon mass is half the double-donut mass.

Unlike Ref.  \cite{BK}, the error bars in Fig.  \ref{fig:glueplot} are
$ \Gamma_{j}, $ the experimental decay widths of each state\cite{pdg}. 
Masses are arguably not known to better accuracy than the widths. 
Theory uncertainties are conservatively estimated using the size of
effects not included, namely the width.  Yet mass parameters can be
fit with great exactness, and BK use\cite{BK} these much smaller
experimental errors.  Meanwhile the central values of Fig. 
\ref{fig:glueplot} are so embarrassingly constant that the error bars
are either overestimated, or something very deep is happening.  In
ordinary data, fluctuations of values would be comparable to the width
of the error bars.  This not seen: the $\chi^{2}/dof $ value of the
data shown is $3 \times 10^{-2} /16$, while it should be about one. 
Fig.  \ref{fig:glueplot} is not a mistake but an honest mystery.  BK
sidestep this mystery because they use the experimental mass
uncertainties, which are so much more tiny than the widths.  We can
speculate that the true poles of the relevant Green functions in the
complex energy plane are entirely set by topological rules,
reminiscent of the Veneziano model \cite{venitian}, while the decay to
ordinary hadrons is just unrelated messiness.  Other puzzles can be
mentioned: rigid classical knots transform like tensors, which is spin
$J=2$.  Meanwhile BK find $J$=0, 1, 2, 4 states.  Where are the
stringy excitations (vibrational modes) of the knots?  There are right
and left-handed trefoils, and many other knots, making parity $P=+, \,
-$ (even and odd ) combinations.  Yet only $P=+$ is seen in the
data.  An ``even parity'' rule is needed, which happens to be a
feature of low-derivative invariants in our theory.  Whether other
states exist, or why the topological parity does not contribute is
unknown.  All states have even charge conjugation $C=+$ , which is
also consistent with the low-derivative invariants.
 
The evidence of the knotted glueballs indicates that an $SO(2) \sim
U(1)$ subgroup of the fundamental local symmetry may penetrate all the
way into the effective theory.  There is a hope that a broad stream of
phenomenology, from the flux tubes of Regge theory to those invoked in
quark confinement, might have their justification and unification via
simple observations on the length of knotted rope.

\end{document}